\documentclass[preprint,onecolumn,floatfix,superscriptaddress,amsmath,amsfonts,amssymb,aps]{revtex4}

\usepackage{graphicx}
\usepackage{dcolumn}
\usepackage{bm}
\usepackage[normalem]{ulem}
\usepackage{color}
\usepackage{amssymb}
\usepackage{amstext}

\begin{document}


\title{Morphology and Optical Properties of Thin \\ 
Cd$_3$As$_2$ Films of a Dirac Semimetal Compound}

\author{Natalia Kovaleva}
\email{kovalevann@lebedev.ru}
\affiliation{Lebedev Physical Institute, Russian Academy of Sciences\\
\hspace{0.46cm} Leninsky prospect 53, Moscow 119991, Russia}
\author{Ladislav Fekete}
\author{Dagmar Chvostova}
\affiliation{Institute of Physics, Academy of Sciences of the Czech Republic\\
\hspace{0.46cm} Na Slovance 2, Prague 18221, Czech Republic}
\author{Andrei Muratov}
\affiliation{Lebedev Physical Institute, Russian Academy of Sciences\\
\hspace{0.46cm} Leninsky prospect 53, Moscow 119991, Russia}

\date{\today}

\begin{abstract}
Using atomic-force microscopy (AFM) and wide-band (0.02--8.5\,eV) spectroscopic ellipsometry techniques we investigated morphology and 
optical properties of Cd$_3$As$_2$ films grown by non-reactive rf magnetron sputtering on two types of oriented crystalline substrates (100)$p$-Si and (001) $\alpha$-Al$_2$O$_3$. The AFM study revealed grainy morphology of the films due to island incorporation during the film growth. The complex dielectric function spectra of the annealed Cd$_3$As$_2$/Al$_2$O$_3$ films manifest pronounced interband optical transitions at 1.2 and 3.0\,eV, in excellent agreement with the theoretical calculations for the body centered tetragonal Cd$_3$As$_2$ crystal structure. We discovered that due to electronic excitations to the Cd(s) conical bands the low-energy absorption edge of the annealed Cd$_3$As$_2$ films reveals linear dependence. We found that for the annealed Cd$_3$As$_2$ films the Cd(s) conical node may be shifted in energy by about 0.08--0.18\,eV above the heavy-flat As(p) valence band, determining the optical gap value. The as-grown Cd$_3$As$_2$ films exhibit the pronounced changes of the electronic band structure due to doping effect associated with Cd non-stoichiometry, where fine-tuning of Cd concentration may result in the gapless electronic band structure of Dirac semimetals.
\end{abstract}

\pacs{Valid PACS appear here}
\maketitle
\section{Introduction}
The Dirac semimetals (DSM) considered as a 3D analog of graphene 
have recently attracted an exceptional attention as materials 
with fundamentally new electronic properties \cite{Wang,Armitage}, 
which may result in a breakthrough for potential applications of 
the next generation of electronic devices. A key feature of the DSM 
is their inverted electronic band structure characterized by the symmetry-protected band crossing the Brillouin zone at the Fermi level or close to the Fermi level. In the proximity to these symmetry-protected nodal points (Dirac points), the gapless electronic excitations reveal linear dispersion with 
rigidly coupled spin and momentum vectors leading to double 
chiral degeneracy of the bands. Naturally, the peculiar features of 
DSM are most pronounced if the Fermi level is located near the 
Dirac points, where the doping effect may provide an insight 
into the topological phases and can be constructive in finding 
new properties for their applications.

Among several candidates, tetragonal $\alpha$-Cd$_3$As$_2$ 
is considered to be one of the most promising to host the Dirac 
semimetal phase in which charge carriers are Dirac fermions with the highest carrier mobility up to 10 $^2$/(V$\cdot$s), which have zero effective mass. Here, the Dirac points for gapless excitations appear at 
two points in the momentum space at $k_z$\,=\,$\pm k_0$ on the $k_z$ axis. The four-fold rotation ($C_4$) symmetry around the $z$-axis forbids the gap-opening. However, the orbital mixing is possible in $x$ or $y$ directions, where 
the Fermi surfaces around the Dirac points can naturally exhibit an 
orbital mixing and Cooper pairing between different orbitals may be 
possible. According to the theoretical predictions, topological superconductivity (SC) related to the Majorana zero modes \cite{Sato} may be realized in DSM under high-pressure conditions or due to doping effects. However, 
the accomplishment of topological SC is still under intense debate, since it is necessary to exclude various effects related to the possible 
symmetry lowering due to crystal deformation accompanying 
structural transitions under the applied pressure \cite{He}. 
In the recent study, SC was discovered in the magnetron sputtered 
Cd$_3$As$_2$ polycrystalline films \cite{Kochura}, 
which revealed several features, such as the presence of tetragonal 
crystalline phase and Shubnikov - de Haas oscillations observed in 
high magnetic fields, indicating its possible topological nature \cite{Suslov}.

In the present study, thin Cd$_3$As$_2$ films grown by non-reactive rf magnetron sputtering in an argon atmosphere on oriented single-crystalline Si (100) $p$-Si and (001) $\alpha$-Al$_2$O$_3$ substrates were investigated by atomic force microscopy (AFM) and wide-range (0.02--8.5\,eV) spectroscopic ellipsometry (SE). The AFM study of the Cd$_3$As$_2$ films implies that the films are continuous and have a granular surface structure. 
We investigated a set of the as-grown and annealed Cd$_3$As$_2$ films and analyzed the effect of cadmium nonstoichiometry on the optical properties of the films. The results of the present study can be useful in the comprehension of the doping effect associated with cadmium non-stoichiometry on the electronic band structure of thin Cd$_3$As$_2$ films and constructive in selecting the films where the new properties related to 3D Dirac fermions including SC may be found.    

\section{Materials and Methods}
Polymorphism is a peculiar feature of the Cd$_3$As$_2$ compound, which may crystallize in four modifications $\alpha$ ($I4_1cd$), $\alpha'$ ($P4_2/nbc$), $\alpha''$ ($P4_2/nmc$), and $\beta$ ($P4_232$). The structural phase 
transitions occur as follows $\alpha\rightarrow 503\,{\rm K} \rightarrow \alpha' \rightarrow 738\,{\rm K} \rightarrow \alpha" \rightarrow 868\,{\rm K} \rightarrow \beta$.
Thin Cd$_3$As$_2$ films investigated in the present study were grown by non-reactive rf magnetron sputtering in an argon atmosphere on polished oriented single-crystalline (100)$p$-Si and (001) $\alpha$-Al$_2$O$_3$ wafers (for more detail, see Ref.\,\cite{Kochura}). Since the (0001) $\alpha$-Al$_2$O$_3$ and (224) $\alpha$-Cd$_3$As$_2$ have the similar structure and their interatomic (interstitial) distances differ by only 6\%, the oriented film growth can be promoted. However, the conditions for the Si substrates differ, where there is no any structural match between Si and Cd$_3$As$_2$, which is also aggravated by the presence of the amorphous native oxide layer at the surface of the used Si substrates. The films were prepared without heating the substrates, 
with heating the substrates to 520\,K, and with annealing the films at 520\,K in an argon atmosphere. 

The composition of the grown films was nearly stoichiometric, as 
it was verified by energy dispersive X-ray analysis, 
demonstrating that the actual elemental composition is close to the 
stoichiometric Cd$_3$As$_2$ within 2\% accuracy \cite{Suslov}. 
The phase composition of the grown Cd$_3$As$_2$ films was characterized by X-ray diffraction analysis (XRD) at room temperature \cite{Kochura}. The XRD data for the polycrystalline Cd$_3$As$_2$ films synthesized by magnetron sputtering \cite{Suslov} indicate the presence of both $\alpha$ (space group $I4_1/acd$m, $a$=12.6461 \AA, $c$=25.4378 \AA) and $\alpha'$ (space group $P4_2/nbc$, $a$=12.6848, $c$=25.4887 \AA) phases. It was found that the annealing and film growth on the heated substrate led to an increase in the degree of crystallinity and a decrease in the lattice parameters, which is especially pronounced for the Si substrates. 

The surface morphology of the Cd$_3$As$_2$ films grown by
rf sputtering deposition on the polished oriented single-crystalline (100)$p$-Si and (001) $\alpha$-Al$_2$O$_3$ wafers investigated in the present study was characterized by atomic force microscopy (AFM) using an ambient AFM (Bruker, Dimension Icon) in Peak Force Tapping mode with ScanAsyst Air tips (Bruker; k=0.4\,N/m; nominal tip radius 2\,nm). 

Optical properties (complex dielectric function and refractive coefficient spectra, optical conductivity and optical absorption spectra) of the Cd$_3$As$_2$/Si and Cd$_3$As$_2$/Al$_2$O$_3$ films were investigated in the wide photon energy range 0.02\,--\,8.5\,eV with a set of three J.A. Woollam spectroscopic ellispometers: IR-VASE Mark II, VASE, and VUV-Gen II. The ellipsometry measurements were performed at several angles of incidence at room temperature. At each angle of incidence, the raw experimental data are represented by real values of the ellipsometric angles $\Psi(\omega)$ and $\Delta(\omega)$. These values are defined through the complex Fresnel reflection coefficients for light-polarized parallel $r_p$ and perpendicular $r_s$ to the plane of incidence as follows
${\rm tan}\,\Psi\,e^{i\Delta}=\frac{r_p}{r_s}$.
The measured ellipsometric angles, $\Psi(\omega)$ and $\Delta(\omega)$, were simulated using multilayer models available in the J.A. Woollam VASE software \cite{VASE}. In the simulation, the complex dielectric function spectra of the blank Si and Al$_2$O$_3$ substrates were represented by their tabular dielectric function spectra \cite{Palik}. 

\section{Results}

\subsection{Atomic force microscopy study of Cd$_3$As$_2$ films grown by non-reactive rf magnetron sputtering}
\begin{figure}
\centering
\includegraphics[width=10.0cm]{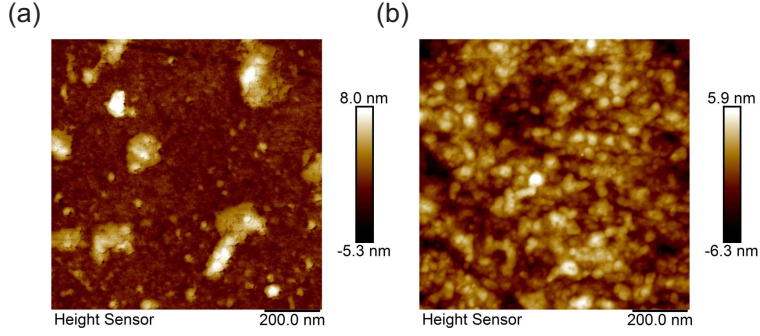}
\caption{Typical AFM topography profiles of the Cd$_3$As$_2$ films grown by non-reactive rf magnetron sputtering on the polished oriented single-crystalline wafers (a) (100)$p$-Si and (b) (001) $\alpha$-Al$_2$O$_3$. The scan size is 1\,$\times$\,1 $\mu$m$^2$.}
\label{AFM}
\end{figure}

Figure\,\ref{AFM}(a,b) shows typical 1\,$\times$\,1 $\mu$m$^2$ AFM images characterizing the surface morphology structure of the annealed Cd$_3$As$_2$ films grown by non-reactive rf magnetron sputtering on the polished oriented single-crystalline (100)$p$-Si and (001) $\alpha$-Al$_2$O$_3$ wafers, respectively. The surface morphology of the Cd$_3$As$_2$/Si film sample (Fig.\,\ref{AFM}(a)) exhibits prominent features represented by isolated 100--200\,nm-sized equiaxed grains, which appear in bright contrast in the AFM image while scanning the film surface. The peculiar surface morphology was reported earlier for 
the magnetron-sputtered Cd$_3$As$_2$ films, where it was also attested that this granular structure was not affected by annealing \cite{Kochura,Suslov}. The recorded AFM image (Fig.\,\ref{AFM}(a)) shows that the surface regions between large isolated by about 100--400\,nm grains, include smaller 10--30\,nm-sized grains, and even still smaller grains. These large and small protruding nanoisland structures seemingly characterize the initial stage of the Cd$_3$As$_2$ film growth on the (100)$p$-Si substrates determined by the rf sputtering conditions promoting island growth. The island height is less than 10\,nm, being about several times smaller than the film thickness, and the grown Cd$_3$As$_2$/Si films of about 100\,nm thick are continuous. We would like to note that the scanning electron microscopy images of the Cd$_3$As$_2$ films exhibiting the subkelvin superconductivity without any external stimuli \cite{Suslov} revealed a similar surface morphology as the grainy surface structure of the Cd$_3$As$_2$/Si film shown in Fig.\,\ref{AFM}(a). By contrast, the surface morphology of the Cd$_3$As$_2$/Al$_2$O$_3$ film sample is different as demonstrated by the AFM image shown by Fig.\,\ref{AFM}(b). One can see that here the grainy morphology structure of the film is represented by relatively small grains having the size of about 10--30\,nm, where the grains are closely situated to each other. Seemingly, these small nanoislands characterize 
the initial stage of the Cd$_3$As$_2$ film growth on the (001) $\alpha$-Al$_2$O$_3$ substrates determined by the rf sputtering and [112] texturing conditions. The island height is less than 10\,nm, so relatively thick Cd$_3$As$_2$/Al$_2$O$_3$ films with thickness several times higher than the island height can be regarded as continuous.    

\subsection{Spectroscopic ellipsometry study of the Cd$_3$As$_2$ films}
Complex dielectric function spectra, $\tilde\varepsilon(\omega)=\varepsilon_1(\omega)+{\rm i}\varepsilon_2(\omega)$, of the investigated Cd$_3$As$_2$/Si and Cd$_3$As$_2$/Al$_2$O$_3$ film samples were probed in the wide photon energy range 0.02\,--\,8.5\,eV with a set of three J.A. Woollam spectroscopic ellispometers: IR-VASE Mark II, VASE, and VUV-Gen II. The ellipsometric angles $\Psi(\omega)$ and $\Delta(\omega)$ were measured at two or three angles of incidence of 60$^\circ$, 
65$^\circ$, and 70$^\circ$ at room temperature.   
The obtained ellipsometric angles $\Psi(\omega)$ and $\Delta(\omega)$ were simulated in the framework of the bilayer model for the Cd$_3$As$_2$/Si and Cd$_3$As$_2$/Al$_2$O$_3$ film samples using the J.A. Woollam VASE software \cite{VASE} following the same approach as presented in more detail in our recent study of metallic Ta films \cite{Kovaleva_metals}.

The complex dielectric function $\tilde \varepsilon(\omega)=\varepsilon_1(\omega)+{\rm i} \varepsilon_2(\omega)$ of a Cd$_3$As$_2$ layer was modeled by the multiple Gaussian functions 
\begin{eqnarray}
\tilde\varepsilon(E\equiv \hbar\omega)=\varepsilon_{\infty}+\sum_n \left( \label{DispAna}
\tilde \varepsilon_{Gauss} \right)_n=\varepsilon_{\infty}+\sum_n \left( \varepsilon_{1_n} + {\rm i} \varepsilon_{2_n} \right),\\ 
\varepsilon_{2_n}=A_n e^{-\left( \frac{E-E_n}{\sigma}\right)^2}-A_n e^{-\left( \frac{E+E_n}{\sigma}\right)^2},\\
\sigma=\frac{\gamma_n}{2\sqrt{ln(2)}},
\end{eqnarray}
where $\varepsilon_{\infty}$ is the high-frequency dielectric constant, which takes into account the contribution of the high-energy interband transitions. The adjustable Gaussian parameters were $E_n$, $\gamma_n$, and $A_n$ of the peak energy, the half width at half maximum, and the $\varepsilon_2$ peak height, respectively.

\begin{figure}
\centering
\includegraphics[width=11.0cm]{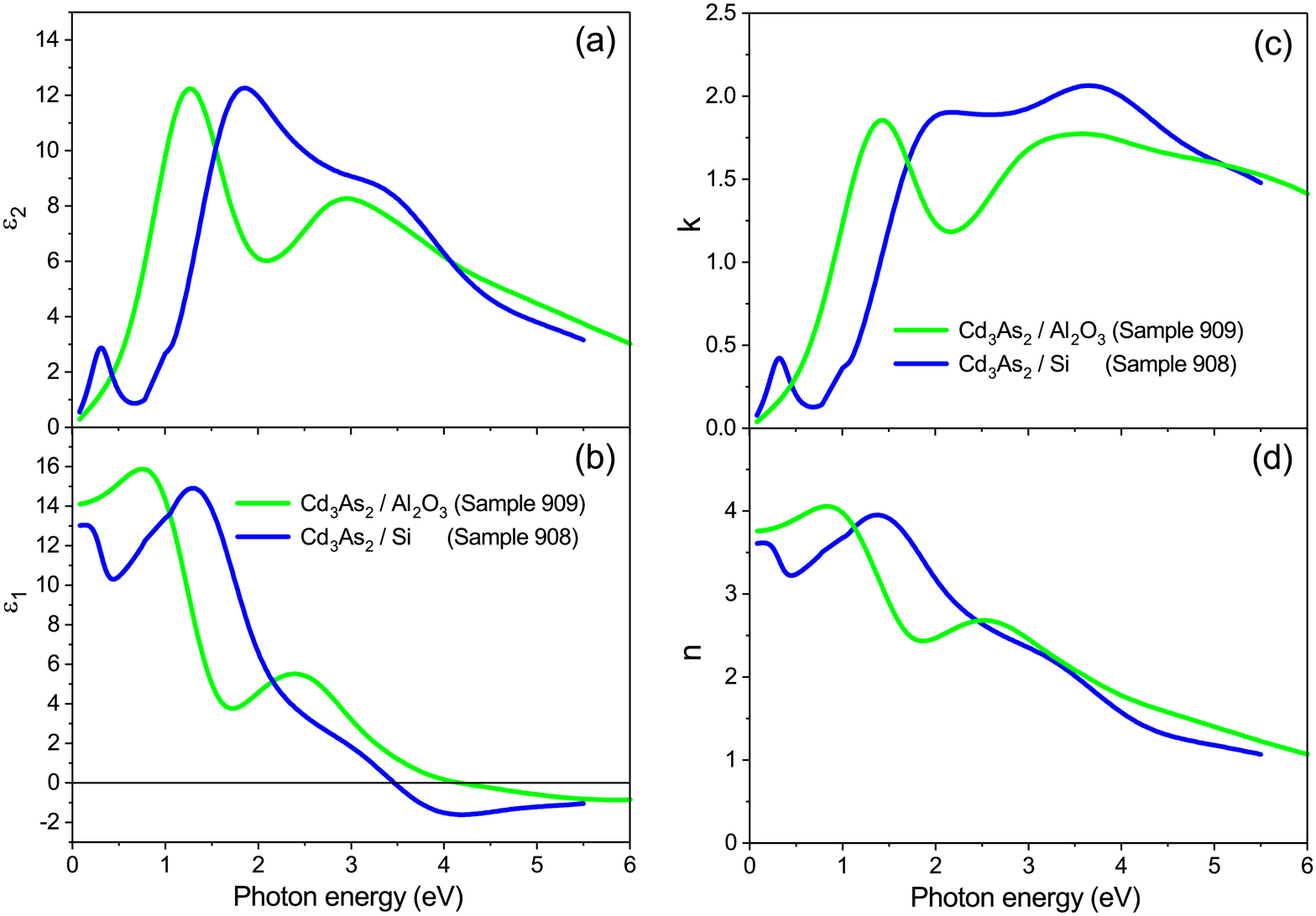}
\caption{(a,c) The imaginary parts of the dielectric function $\varepsilon_2(\omega)$ and refractive index $k$ and (b,d) the real parts of the dielectric function $\varepsilon_1(\omega)$ and refractive index $n$ for the annealed Cd$_3$As$_2$/Si (sample 908, 10\,W, 80\,min, 20\,min annealing at 520\,K) and Cd$_3$As$_2$/Al$_2$O$_3$ (sample 909, 10\,W, 40\,min, 20\,min annealing at 520\,K) films, shown by solid blue and green curves, respectively. The thickness estimated from the model simulations was 91$\pm$5\,nm for the Cd$_3$As$_2$/Si film and 46$\pm$3\,nm for the Cd$_3$As$_2$/Al$_2$O$_3$ film.}
\label{Cd3As2epsnk}
\end{figure}

\begin{figure}
\centering
\includegraphics[width=7.0cm]{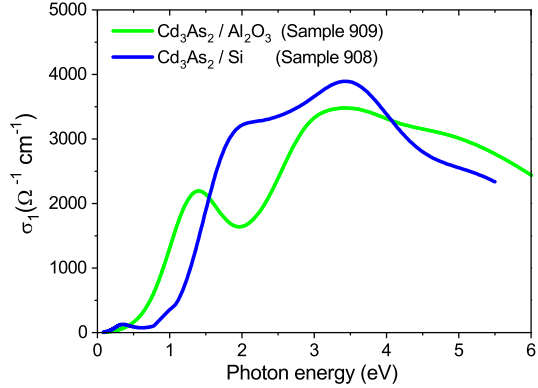}
\caption{The respective optical conductivity spectra, $\sigma_1(\omega)=\frac{1}{4\pi}\omega\varepsilon_2(\omega)$, for the annealed Cd$_3$As$_2$ films.}
\label{Cd3As2_sigma}
\end{figure}

\begin{figure}
\centering
\includegraphics[width=13.0cm]{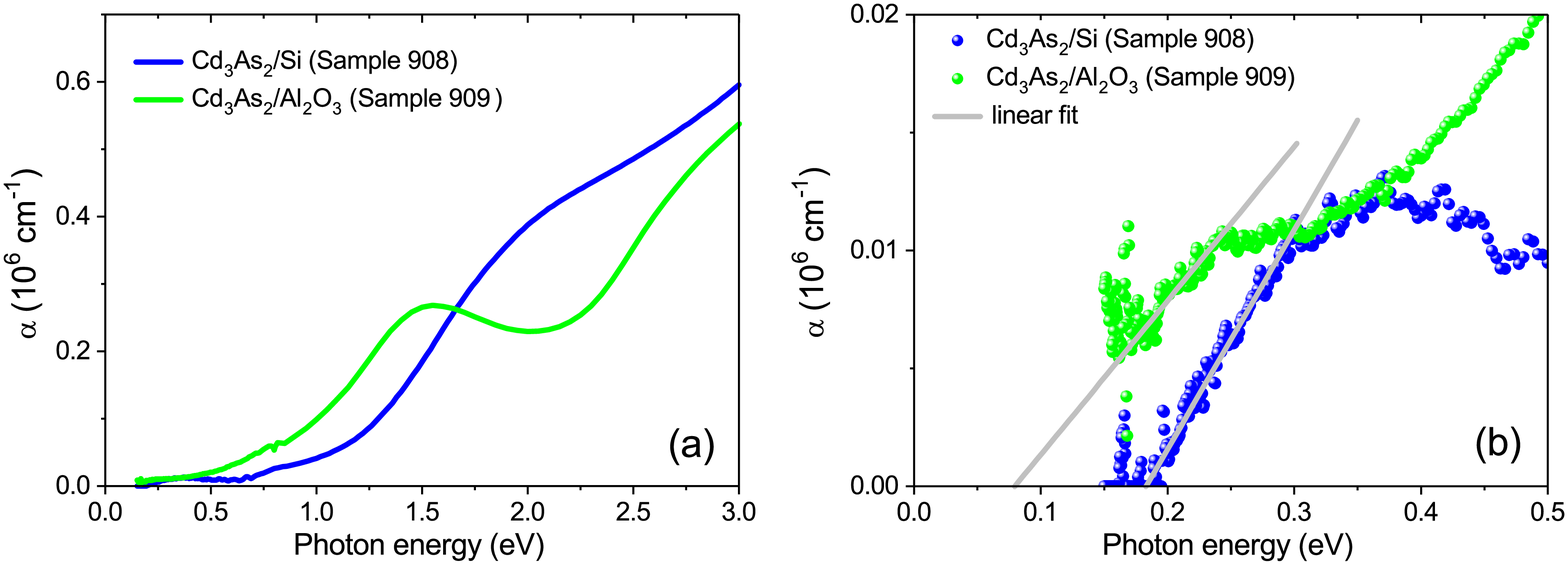}
\caption{The absorption spectra $\alpha(\omega)$ for the annealed Cd$_3$As$_2$/Si and Cd$_3$As$_2$/Al$_2$O$_3$ films shown for (a) the wide spectral range and (b) at the low-energy edge (shown by solid blue and green curves and symbols, respectively). 
The gray solid lines show linear extrapolations ($y=a+bx$) of the low-energy absorption edge (for Cd$_3$As$_2$/Si: $a=-0.017\pm2\%$, $b=0.093\pm2\%$ and for Cd$_3$As$_2$/Al$_2$O$_3$: $a=-0.005\pm9\%$, $b=0.065\pm4\%$).}
\label{abs}
\end{figure}

In the simulation of the ellipsometric angles, $\Psi(\omega)$ and $\Delta(\omega)$, the Cd$_3$As$_2$ layers at the Si and Al$_2$O$_3$ substrates were described by different multiple Gaussian dispersion models [Eq.\,(\ref{DispAna})].
To the utilized multiple Gaussian model, the complex dielectric function spectra of the blank Si and Al$_2$O$_3$ substrates were substituted by the tabular complex dielectric function spectra \cite{Palik}. The quality of the fit for the studied Cd$_3$As$_2$/Si and Cd$_3$As$_2$/Al$_2$O$_3$ film samples was verified by the coincidence with the recorded ellipsometric angles $\Psi(\omega)$ and $\Delta(\omega)$ within the specified accuracy lower than 5\%. The good quality of the fit allowed us to estimate an actual film thickness of the films under study. From the multiple Gaussian model simulations by using Eq.(\ref{DispAna}), the imaginary and real parts of the complex dielectric function spectra, $\varepsilon_2(\omega)$ and $\varepsilon_1(\omega)$, as well as the imaginary and real parts of the complex refractive index, $k$ and $n$, the optical conductivity and absorption spectra of the Cd$_3$As$_2$ layer on the Si and Al$_2$O$_3$ substrates were obtained as displayed in Figs.\,\ref{Cd3As2epsnk}--\ref{Else_e1e2}.

From Fig.\,\ref{Cd3As2epsnk}(a,b) one can monitor the properties of the 
complex dielectric function spectra for the annealed Cd$_3$As$_2$ films grown by non-reactive rf magnetron sputtering on the polished oriented single-crystalline substrates (100)$p$-Si and (001) $\alpha$-Al$_2$O$_3$. We found that, on the one hand, the dielectric response of the annealed Cd$_3$As$_2$/Al$_2$O$_3$ film measured in the wide spectral range manifests two clearly pronounced interband optical transitions peaking at around 1.2 and 3.0\,eV. On the other hand, the wide-range complex dielectric response measured for the annealed Cd$_3$As$_2$/Si film looks notably different. Namely, the evidently pronounced peak appears here at low photon energies at $\sim$\,0.36\,eV. In addition, the main interband optical transitions are shifted to higher energies to 1.9 and 3.5\,eV. Figure\,\ref{Cd3As2_sigma} shows the associated optical conductivity spectra, $\sigma_1(\omega)=\frac{1}{4\pi}\omega \varepsilon_2(\omega)$, which follow the main trends observed in their complex dielectric function spectra. Figure\,\ref{abs}(a,b) shows the optical absorption spectra for the annealed Cd$_3$As$_2$ films in the wide-range of the photon energies, where we also give a more detailed absorption at the low-energy edge. We discovered that the low-energy absorption edge of the investigated films exhibits linear dispersion, which is shifted to the higher energies and, therefore, more clearly pronounced for the annealed Cd$_3$As$_2$/Si film. We found that the low-energy absorption edge can be well described by the parameters of linear extrapolation (for more detail, see Fig.\,\ref{abs}(b)). 

Here, we also investigated the as-grown Cd$_3$As$_2$ films prepared by non-reactive rf magnetron sputtering on the Si substrates, which exhibited the peculiar morphology properties shown in Fig.\,\ref{AFM}(a). We would like to note that namely for the as-grown films the subkelvin SC was reported, which occurred without any external stimuli \cite{Suslov}. The optical properties of the as-grown films were probed by using a J.A. Woollam VASE spectroscopic ellipsometer in the spectral range from 0.7 to 6.0\,eV, and the result of the model simulations is presented in Fig.\,\ref{Else_e1e2} (a,b). One can notice from the figure that the as-grown Cd$_3$As$_2$/Si samples show modified optical properties in the whole studied spectral range, as at the low energies, so in the range of the main interband transitions. Here, in comparison to the 
annealed Cd$_3$As$_2$/Si film, the low-energy feature may disappear or shifted to the higher energy of 1\,eV. At the same time, the first interband transition may show a noticeable blue shift to the higher energy of about 2.2\,eV. 

\section{Discussion}
The imaginary part of the dielectric function spectra (see Fig.\,\ref{Cd3As2epsnk}(a)) displaying two main electronic interband transitions at 1.2 and 3.0\,eV for the annealed Cd$_3$As$_2$/Al$_2$O$_3$ film is in a perfect agreement with the theoretical calculations for the body centered tetragonal Cd$_3$As$_2$ crystal by Conte {\em et\,al.} (see Fig.\,10 in Ref.\,\cite{Conte}). 
From these theoretical calculations, no appreciable anisotropy is foreseen for the in-plane $(\varepsilon_{xx}+\varepsilon_{yy})/2$ and out-of-plane $\varepsilon_{zz}$ components. However, the wide-range dielectric response measured for the annealed Cd$_3$As$_2$ film/Si looks notably different. 
Namely, the evidently pronounced peak appears here at low photon energies at $\sim$\,0.36\,eV. In addition, the main interband optical transitions are shifted to the higher energies of 1.9 and 3.5\,eV. We refer the observed difference in the dielectric function spectra to the effect of different doping associated with cadmium non-stoichiometry in the annealed Cd$_3$As$_2$ /Al$_2$O$_3$ and Cd$_3$As$_2$/Si films.

\begin{figure}
\centering
\includegraphics[width=8.5cm]{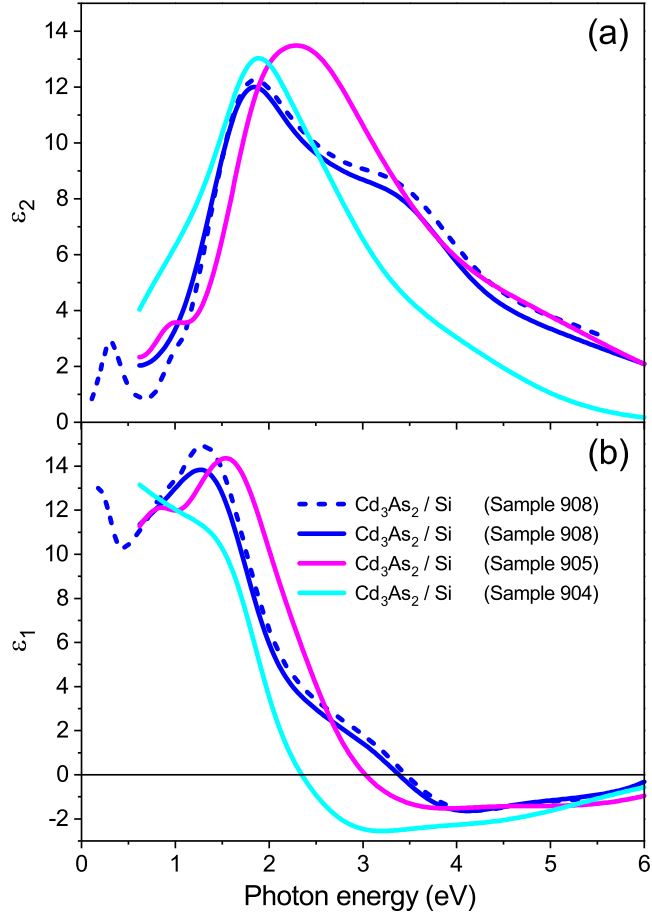}
\caption{(a,b) The imaginary $\varepsilon_2(\omega)$ and real $\varepsilon_1(\omega)$ parts of the complex dielectric function of the as-grown Cd$_3$As$_2$ films on the polished oriented single-crystalline wafers (100)$p$-Si. The dashed blue curve corresponds to the data shown by solid blue line in Fig.\,\ref{Cd3As2epsnk}. The solid blue, cyan, and magenta curves correspond to the ellipsometry measurements for the samples 908 (annealed), 904 (as-grown, 10 W, 20 min), and 905 (as-grown, 10 W, 80 min), respectively, obtained using a J.A. Woollam VASE ellipsometer.}
\label{Else_e1e2}
\end{figure}

In Fig.\,\ref{abs}(b) we present the optical absorption spectra, 
$\alpha(E \equiv \hbar \omega)$, obtained from the modeling of the SE data using a commercial WVASE32 
software package \cite{VASE}. We found that the low-energy absorption edge of the annealed Cd$_3$As$_2$/Si film is better described by the linear dependence at the lowest measured photon energies from 0.18 to 0.3\,eV, 
compared to the commonly applied Tauc-type dependencies for direct, 
$\left( \alpha E \right)^2 \propto \left( E-E_d \right)$, 
and indirect, 
$\left( \alpha E \right)^{1/2} \propto \left( E-E_i \right)$, gaps. 
The observed absorption characterized by the linear dispersion can be associated with the conical absorption. Indeed, due to conical dispersions, the optical absorption of 3D massless Kane particles is proportional to the frequency $\omega$ or to the photon energy ($E\equiv \hbar \omega$) \cite{Akrap,Orlita,Timusk}, distinctly in contrast to frequency-independent optical conductivity of 2D Dirac electrons as observed in graphene \cite{Kuzmenko,Nair,Kovaleva_2D_Materials}. Massless Kane electrons, which are not symmetry protected, may exist in a semiconductor with a nearly vanishing gap. We suggest that due to doping effect the conical Cd(s) conduction band minimum is shifted in energy above the heavy-flat As(p) valence band by about 0.183$\pm$0.007\,eV in the annealed Cd$_3$As$_2$/Si film, determining the optical gap value. Here, the maximum appeared at $\sim$\,0.36\,eV can be associated with the absorption threshold developing near 0.3 eV. This can also reasonably explain the observed blue shift occurring for the main electronic transitions, which appear at 1.9 and 3.5\,eV. 
By contrast, the linear dispersion due to conical absorption in the annealed Cd$_3$As$_2$/Al$_2$O$_3$ seems to appear at the lower energy around 
0.08$\pm$0.01\, eV determining the value of the 
optical gap in accord with the linear extrapolation to zero 
photon energies. We note that because of the lack of the reliable data at low photon energies, the given linear approximation may be not very accurate. Here, the low-energy feature becomes substantially weaken. This also results in smooth crossover from the conical to interband absorption. A variety of other investigated as-grown and annealed Cd$_3$As$_2$ films/Si (see Fig.\,\ref{Else_e1e2}) illustrates that pronounced changes of the electronic band structure take place at low energies due to the conical absorption, as well as in the range of interband transitions due to Cd non-stoichiometry (doping effect). The low- and high-energy trends seems to be in correlation as discussed above. 

Another issue is related to the possibility of observation of 
SC related to 3D Dirac massless fermions, which is 
reported to exist below 300 mK for the as-grown Cd$_3$As$_2$ films prepared by non-reactive rf magnetron sputtering (as those investigated in the present study) by Suslov et al. \cite{Suslov}. In Fig.\,\ref{Else_e1e2}(a,b) we show the complex dielectric function spectra of the as-grown Cd$_3$As$_2$/Si films along with the annealed Cd$_3$As$_2$ film/Si. The dielectric function spectra for the as-grown samples look notably different at low energies, as well as in the range of interband transitions. As we have discussed, this can be explained by the shift in energy of the conical Cd(s) node, which is strongly dependent on the cadmium stiochiometry due to doping effect. The SC properties necessarily disappear in the annealed films under study due to opening of the gap, which can be attested as being beyond the playground for finding there new SC properties associated with 3D Dirac fermions. Then, one might suggest that a slight deviation from the intrinsic stoichiometry peculiar to Cd$_3$As$_2$ single crystals leading to the appearance of the conical bands very close to the heavy-flat As(p) valence band around the $\Gamma$ point in the Brillouin zone may be the necessary but certainly not sufficient condition for the search for the new SC properties due to 3D Dirac massless fermions in Cd$_3$As$_2$ films. 
Indeed, the symmetry-protected 3D Dirac particles, if existing in Cd$_3$As$_2$, may appear in a very short energy scale given by the crystal field splitting (at most a few tens meV). 

\section{Conclusions}
In summary, here using atomic force microscopy and wide-range (0.02-8.5\,eV) spectroscopic ellipsometry we have studied morphology and optical properies of the as-grown and annealed Cd$_3$As$_2$/Si and Cd$_3$As$_2$/Al$_2$O$_3$ films prepared by rf magnetron sputtering. The AFM study of the Cd3As2 films implies that the films are continuous and have a granular structure with island incorporation during the film growth. The complex dielectric function of the annealed Cd$_3$As$_2$/Al$_2$O$_3$ film manifests pronounced interband optical transitions at 1.2 and 3.0\,eV, in excellent agreement with the theoretical calculations for the body centered tetragonal Cd$_3$As$_2$ crystal by Conte {\em et\,al.} \cite{Conte}. The dielectric function response for the annealed Cd$_3$As$_2$/Si film looks 
notably different, where the evidently pronounced peak appears at 
low photon energies at $\sim$\,0.36\,eV. We found that the absorption edge near the low-energy feature exhibits a linear dependence and can be associated with the conical absorption. The Cd(s) conical node is shifted in energy above the heavy-flat As(p) valence band by about 0.183$\pm$0.007\,eV, determining the optical gap value. The as-grown Cd$_3$As$_2$/Si films exhibit the pronounced changes of the electronic band structure due to Cd non-stoichimetry (doping effect), where the low-energy feature may disappear, signalling gapless electronic band structure. In principal, the conical dispersion at the point of the semiconductor to semimetal topological transition can be achieved by fine-tuning of cadmium concentration. However, since the symmetry-protected 3D Dirac particles may appear in a very short energy scale, insufficient sample quality (inhomogeneous chemical composition and high unintensional doping) might severely complicate this task.\\

{\bf Acknowledgement}

We thank Kochura for providing us with the Cd$_3$As$_2$ film samples. 
We thank Yu. A. Aleshchenko for useful discussions. The work of N.N.K. is carried out within the state assignment of the Ministry of Science and Higher Education of the Russian Federation (theme ``Physics of condensed matter: new materials, molecular and solid state structures for nanophotonics, nanoelectronics, and spintronics'').\\  
 
{\bf Declaration of competing interest}

The authors declare no conflict of interest.

\end{document}